# The Impact of Designated Market Makers on Market Liquidity and Competition:

## A Simulation Approach

BY

Cong Zhou

NYU Stern

April 2023

Professor Mary Billings             Professor Marti G. Subrahmanyam

Faculty Advisor                     Thesis Advisor

# Abstract

This paper conducts an empirical investigation into the effects of Designated Market Makers (DMMs) on key market quality indicators, such as liquidity, bid-ask spreads, and order fulfillment ratios. Through agent-based simulations, this study explores the impact of varying competition levels and incentive structures among DMMs on market dynamics. It aims to demonstrate that DMMs are crucial for enhancing market liquidity and stabilizing price spreads, thereby affirming their essential role in promoting market efficiency. Our findings confirm the impact of the number of Designated Market Makers (DMMs) and asset diversity on market liquidity. The result also suggests that an optimal level of competition among DMMs can maximize liquidity benefits while minimizing negative impacts on price discovery. Additionally, the research indicates that the benefits of increased number of DMMs diminish beyond a certain threshold, implying that excessive incentives may not further improve market quality metrics.

# Contents

# Background

Market liquidity, which allows for the rapid buying and selling of securities without substantially affecting their price, is crucial for the efficient functioning of financial markets. However, during periods of market turbulence, liquidity can suddenly evaporate, as demonstrated by events such as the 2010 Flash Crash. Designated Market Makers (DMMs), primarily high-frequency trading firms, play a pivotal role in providing day-to-day liquidity and are obligated to maintain orderly market conditions.

Numerous studies have underscored the critical role of market liquidity for the functionality and resilience of financial markets, with Designated Market Makers (DMMs) playing a central role in maintaining this liquidity. In the study 'Designated Market Makers: Competition and Incentives,' Professor Subrahmanyam argues that competition among DMMs significantly reduces both quoted and effective spreads, primarily by mitigating adverse selection costs, thus enhancing overall market liquidity. Similarly, Tse & Zabotina, in their paper 'Do Designated Market Makers Improve Liquidity in Open-Outcry Futures Markets?', emphasize the crucial role of DMMs in futures markets, noting significant improvements in liquidity after introducing market makers to the Chicago Board of Trade's 10-year interest rate swap futures contract. Additionally, in 'How Do Designated Market Makers Create Value for Small-Caps?', Menkveld & Wang demonstrate that DMM engagement not only boosts liquidity and reduces liquidity risk but also generates average abnormal returns, highlighting the substantial value DMMs add to the markets, particularly for smaller companies.

Nonetheless, the predominant methodology used in existing research, which analyzes historical market orders to assess the presence and impact of DMMs, overlooks certain subtleties. Specifically, the effects of incremental changes, such as adding a DMM or slightly increasing rebate rates, on market liquidity are not fully captured, as real markets may not undergo these precise adjustments. A simulation-based approach offers a remedy to this limitation, enabling an in-depth exploration of the consequences of altering the number of DMMs and adjusting rebates and fees. Such simulations can illuminate the nuances of price discovery, market liquidity, and equity, thereby providing a more comprehensive understanding of the impact DMMs have on financial markets.

Therefore, I propose developing an agent-based simulation platform for a limit order book market, featuring configurable strategies for market maker agents and adjustable market structures. This platform will allow for direct control and precise measurement, enabling the quantification of the causal impacts of parameter changes in ways that are not possible with empirical data alone. The experiments will systematically vary factors such as the number of DMMs, rebate rates and asset types. The effects of these variations will be assessed through metrics such as bid-ask spreads, market fulfillment ratios, and price deviations from fundamental values.

This research aims to validate existing findings on the roles of Designated Market Makers

(DMMs) and to further explore the effects of incentives provided to DMMs and the competitive dynamics among them on market quality metrics such as liquidity, volatility, spreads, and price efficiency. I hypothesize that beyond a certain threshold, additional incentives or increased competition among DMMs will no longer significantly affect these metrics. Furthermore, I propose that there exists an optimal level of competition among DMMs that effectively balances liquidity provision with minimal market distortions. Additionally, I anticipate that a higher number of DMMs in the market will enhance market efficiency, particularly in stabilizing prices more quickly following market shocks, as more DMMs should help the market's price reflection align more closely with true values. On the other hand, more DMMs may enhance too much competition and harm each individual DMM's P/L performance.

This research has the potential to offer exchanges and regulators actionable insights for designing optimal market-making schemes. By elucidating the impact of various policy levers—such as incentives, competition, entry barriers, and liquidity differentiation—on market quality, it enables informed decision-making to enhance liquidity and stability during turbulent periods. Additionally, the findings from this study may be applicable to broader contexts, including over-the-counter markets, thereby extending its relevance and utility.

# Methodology

I have developed an agent-based model of a limit order book market in Python, which is publicly available on GitHub. The model features several key types of agents:

- Fundamental traders: These agents submit orders based on stochastic signals related to the fundamental value of assets.
- Special participants: These are varied participants each with unique objectives.
- Designated market makers (DMMs): These are enhanced market-making agents equipped with specialized incentives and requirements.

The model's limit order book matches incoming market orders with resting limit orders, generating price time series and order flow data. To examine the impact of DMMs, I will systematically vary parameters such as:

- Number of DMMs (ranging from 1 to 5)
- Maker rebates for DMMs
- Maximum inventory limits
- Maximum quote size (set at 1/5 of maximum inventory)
- DMM quote aggressiveness
- Fundamental value change pattern (initial price and volatility)
- Spread of noise trades
- Annual return of the trading asset
- Annual volatility of the trading asset

The dependent variables to be analyzed include:

- Bid-ask spread
- Depth at the best quote
- Price volatility
- Liquidity during volatility shocks
- Price deviation from fundamental value
- Number of timestamps required for price correction after shocks

Core experiments will involve:

- Varying the number of DMMs at different incentive levels
- Applying volatility shocks to test liquidity resilience
- Introducing informed trading to assess price efficiency

We will both simulate the basic market moves both by importing the real-world historical trade book and simulating the price moves. The simulation will be based on the following distribution:

- Order' time: Poisson Distribution

$$f(x) \ = \ P(X = x) \ = \ (e^{-\lambda}\,\lambda^{x}\,)/x!$$

- Asset fundamental value movement: Brownian Motion

$$\delta P = e^{\left(\mu - \frac{\sigma^2}{2}\right)*\frac{t}{n} + \sigma * X \sim N(0, \sqrt{t/n})}$$

- Price spread: Normal Distribution

- Order quantity: Normal Distribution

The liquidity is measured in various metrics, which includes:
- E/P Ratio – executed to pending orders ratio

- QS – quotes spread computed as the time-weighted average, over $\triangle$t, of $(a_t - b_t)/m_t$, where $a_t$ is the best ask, $b_t$ is the best bid, and $m_t$ the midprice at time t.

- DB – depth at the best quote calculated as the time-weighted average of the sum of the total value of orders resting on the LOB at the best bid and ask.

- RS5 – 5-min realized spread computed as the value-weighted average of
$$q_t(p_t - m_{t+5min})/m_t$$

- RS10 –10-min realized spread computed as the value-weighted average of
$$q_t(p_t - m_{t+10min})/m_t$$

- AS5 – 5min adverse selection computed as the value-weighted average of
$$q_t(m_{t+5min} - m_t)/m_t$$

- AS10 – 10min adverse selection computed as the value-weighted average of
$$q_t(m_{t+10min} - m_t)/m_t$$

- Timestamps till price correction after shocks

To assess the differences in market liquidity across various trading assets, we chose stocks from three publicly traded companies: Coca-Cola (NYSE: KO), Starbucks (Nasdaq: SBUX), and Nvidia (Nasdaq: NVDA). These companies exhibited significant differences in returns and volatility over the past year. By inputting these two variables, we aim to simulate the movements in the fundamental values of trading assets and explore variations in market liquidity.

| Stock | Annual Return | Annual Volatility |
|-------|---------------|-------------------|
| KO | 0.0236 | 0.13 |
| SBUX | -0.27 | 0.254 |
| NVDA | 2.07 | 0.495 |

## Game of Flow of Orders

In the agent-based market making simulation model, various participants interact in the market, both placing orders and providing liquidity to others. The simulation is structured around a series of discrete time steps, denoted as timestamps 0, 1, 2, ..., t, t+1, representing the continuous flow of time. At each timestamp:

- **Market participants receive updates** about the status of orders they submitted in the previous time step.
- **Participants then make new decisions** based on this information, submit new orders, or adjust their existing orders accordingly.
- **The exchange processes these submissions**, matching asks and bids according to current market conditions.
- **Updated order statuses are then sent out**, providing feedback for the next round of decision-making.

This sequential process facilitates a dynamic simulation environment, where market conditions and participant strategies evolve over time, mimicking the functioning of a real-world financial market.

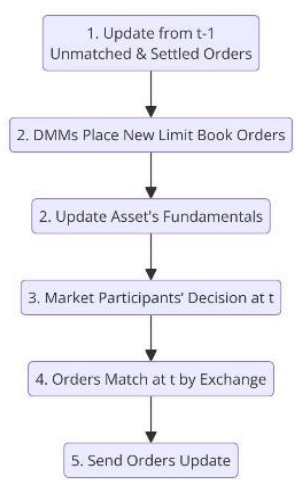

In detail, the game has been cut to numbers of timestamps 0,1,2,3,...., t, t+1, representing the flow of time. For each timestamp, the market participants will receive an update about the orders they submitted at the previous timestamp, make a new decision, and the exchange will begin ask&bid match and send the updated about the orders in the next round.

In the simulation, Designated Market Makers (DMMs) are crucial participants within the market. They have a unique role, making ask and bid decisions with awareness of the orders from other market participants. This process is intentionally streamlined instead of parallel to align with the DMMs' primary objective: to fulfill as many market orders as possible, thus enhancing market liquidity and efficiency. The operational dynamic is akin to that of a physical trading counter where traders walk in to either buy or sell an asset. Traders submit their orders to the DMM, who then assesses the market conditions and sets a price. In the next time step, t+1, traders respond to the price set by the DMM, deciding whether to accept the offer or adjust their order.

Walter Bagehot identified three types of market participants: those with insider information, those wanting to transact assets, and those acting on perceived unrealized information. In this simulation, market participants reflect these categories, each with distinct order logics based on their specific information and objectives.

Participant 1 (P1) represents individuals with insider information. This participant class receives the future price at t+5 already at time t, allowing them to strategize with advanced knowledge. P1 makes trading decisions based on the previous price at t-1 and the future price at t+5, since the price at time t remains unknown until all orders are placed. They are prepared to pay a premium or offer a discount based on their assessment of whether the price is undervalued or overvalued. However, they will refrain from buying or selling if the price difference exceeds 10% from the market price. The general order logistic is shown below:

profit_margin = absolute value of (price(t+5) - price(t-1)) # since the latest known price is at t-1.

**IF** price(t+5) > price(t-1):

    bid_price = price(t-1) + price_preimum * profit_margin # inside trade is willing to pay some premium to secure position

    bid_quantity = min(max(max_inventory - currency_inventory, 0),5) # the maximum quantity in a bid order is 5

    PLACE BUY ORDERS

**ELSE:** # if the price five timestamps later is less than the last price

    ask_price = price(t-1) - price_preimum * profit_margin

    ask_quantity = min(max(max_inventory + currency_inventory, 0),5) # the maximum quantity in a bid order is 5

    PLACE SELL ORDERS

Participant 2 (P2) represents individuals aiming to liquidate their positions. Typically, P2's bid and ask demands are relatively symmetric, making them a key source of market orders. In the simulation, P2 is not viewed as a single individual but as a collective group that consistently places both bid and ask orders at the market price. The general order logistic is shown below:

**IF** there exists a bid&ask order book # Look at bid&ask order book

    **IF** there are open bid limit orders

        Choose the best bid order price and PLACE SELL ORDERS

    **ELSE:**

        **IF** have access to previous matched orders

            Choose the lowest order price that is exercised

                PLACE BUY ORDERS

    **IF** there are open ask limit orders

        Choose the best ask order price and PLACE BID ORDERS

    **ELSE:**

        **IF** have access to previous matched orders

            Choose the highest order price that is exercised

                PLACE SELL ORDERS

Participant 3 (P3) represents individuals who act on outdated information, believing in changes that have already been realized in the market. Specifically, P3 bases their long or short decisions on the price movement from t-2 to t-1, expecting similar trends to occur from t to t+1. Like others, they are willing to offer a premium or a discount based on their expectations.

expected_growth = the asset's fundamental value change from t-2 to t-1, which the player believes to happen at t

profit_margin = absolute value of (price(t-1) - price(t-2))

**IF** price(t-1) > price(t-2):

    bid_price = price(t-1) + price_preimum * profit_margin # Participant 3 is also willing to pay some premium to secure position

    bid_quantity = min(max(max_inventory - currency_inventory, 0),5)

    PLACE BUY ORDERS

**ELSE:**

    ask_price = price(t-1) - price_preimum * profit_margin

    ask_quantity = min(max(max_inventory + currency_inventory, 0),5)

    PLACE SELL ORDERS

Designated market makers in the simulation adjust their decisions based on the orders from P1, P2, and P3. Their primary objective is to facilitate all incoming orders. However, when they detect unusually large orders, which they may attribute to P1, they increase the spread to mitigate potential risks. These market makers also have inventory limits, which prevent them from accepting infinite orders and put pressure on them to manage their holdings effectively. For the calculation of bid and ask prices, we follow the methodology outlined in 'High-frequency trading in a limit order book' by Avellaneda and Stoikov.

The reservation ask price (r_a) is determined as

$$r_a = s + (1 - 2q_t) \times (\sigma^2/2) \times Time\ Horizon_{\square}$$

The reservation bid price(b_a) is determined as

$$r_a = s + (1 - 2q_t) \times (\sigma^2/2) \times Time\ Horizon_{\square}$$

The target weight is calculated as

$$TW = (pmax - s)/(pmax - pmin)$$

S : Current mid-price

q : Dealer's inventory
sigma_squared : 20-Trailing Volatility squared
pmax: 20 – Trailing maximum price
pmin: 20 – Trailing minimum price

## User Interface

The simulation model is equipped with a graphical user interface (GUI). Users can customize various aspects of the simulation through this interface, including setting the number of rounds (which corresponds to the number of timestamps to be observed), the number of market makers, and the price trend setting. For the price trend, users can choose to follow historical market prices or, alternatively, allow the program to generate prices using Brownian motion.

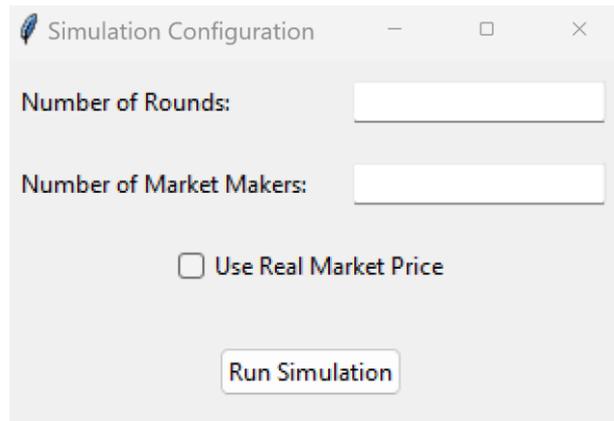

User graphic before simulation begins

More advanced input includes:
- trading assets
- maximum inventory
- starting cash
- price volatility
- price starting point
- price expected return

There are three main outputs from the simulation:
- The cash and inventory report of all participants
- The realized trades across all periods
- All trades status at the end of each timestamp

## Simulation Result

We will generate various market scenarios to assess market liquidity. For each scenario, we will conduct 30 trials, with each trial running through 1,000 timestamps. The simulation will vary the number of designated market makers from 1 to 5, along with a rebate rate ranging from 16 bps to 26 bps, allowing us to explore the relationship between the number of market makers, rebate rate, and market liquidity.

Additionally, we will iterate scenarios with 1 to 15 designated market makers at a fixed rebate rate of 20 bps to analyze price efficiency variations with different numbers of DMMs. Finally, we will re-run scenarios with 1 to 10 DMMs at a 20 bps rebate rate, incorporating shocks to assess market efficiency under volatile conditions.

**Figure 1: Simulation Result of Liquidity in 20 bps rebate rate and 1-5 DMMs**

This series illustrates market liquidity under a fixed rebate rate of 20 bps with a varying number of designated market makers (DMMs) from 1 to 5. Panel A presents a box plot that visualizes the quoted spreads across the five scenarios, showing a decreasing trend as the number of DMMs increases. Panel B charts the depth at the Best Quote (DB), indicating an increase in depth as more DMMs are added. Panel C displays the mean values of RS5, RS10, AS5, and AS10 for each scenario, providing a quantitative view of changes in response size and ask size. The data analyzed is generated through simulation, and the trends suggest that the presence of additional DMMs tends to improve market liquidity, as evidenced by the generally tighter spreads observed.

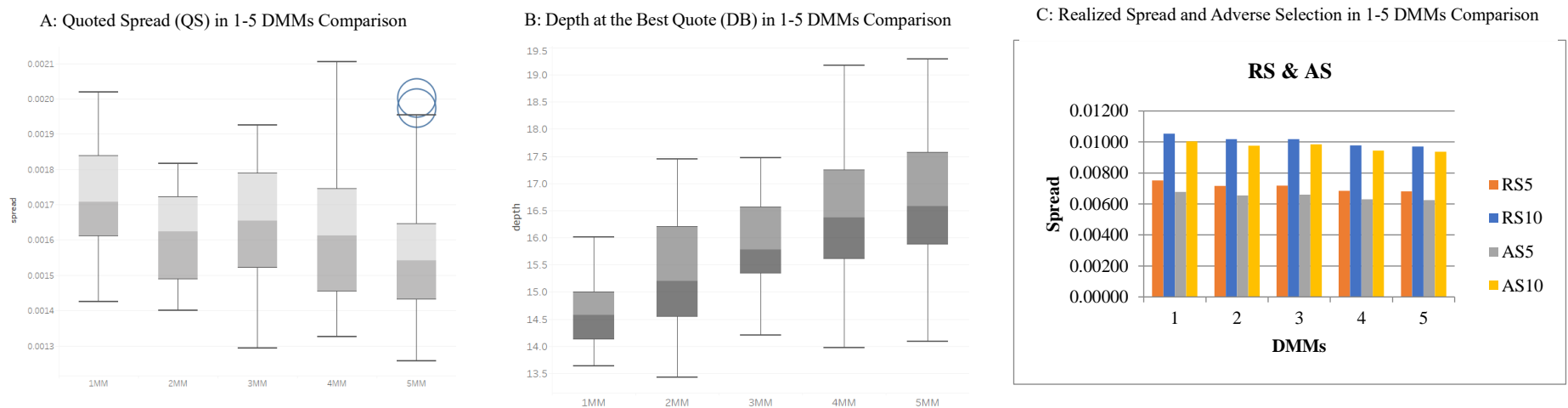

**Figure 2: Simulation Result of Liquidity in 3 DMMs and 16-26 bps Rebate Rate**

This series evaluates market liquidity with three DMMs across rebate rates from 16 to 26 bps. Panel A features a box plot of the EP_ratio, showing an increasing trend as rebate rate increases. Panel B displays the quoted spread for six scenarios, which presents present an unexpected trend: the average quoted spread (QS) decreases as the rebate rate decreases. Panel C shows no statistically significant differences in depth at the Best Quote (DB) as rebate rates change. Panel D presents slight fluctuations of values of RS5, RS10, AS5, and AS10. All data is generated through simulation.

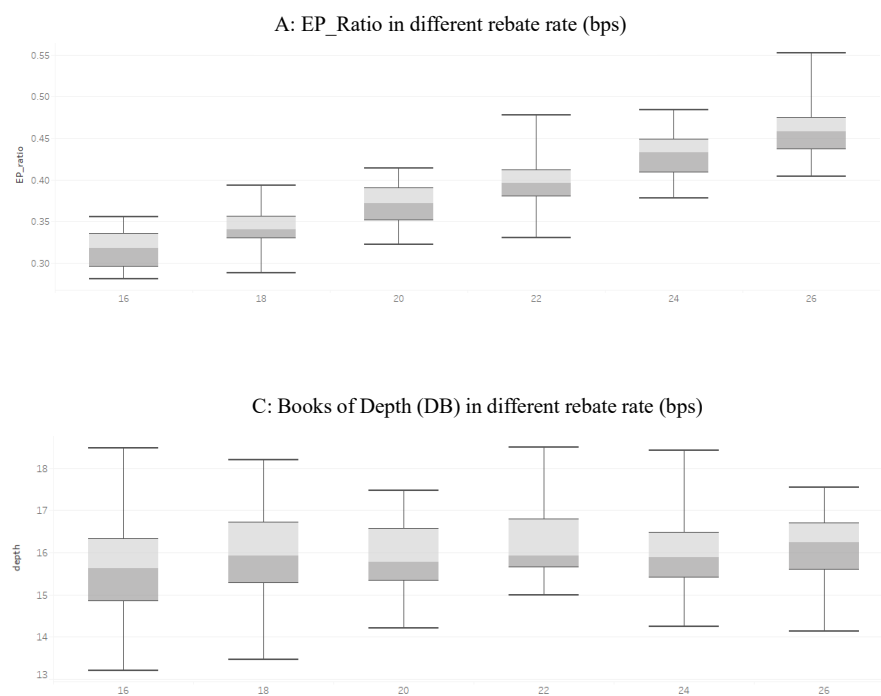

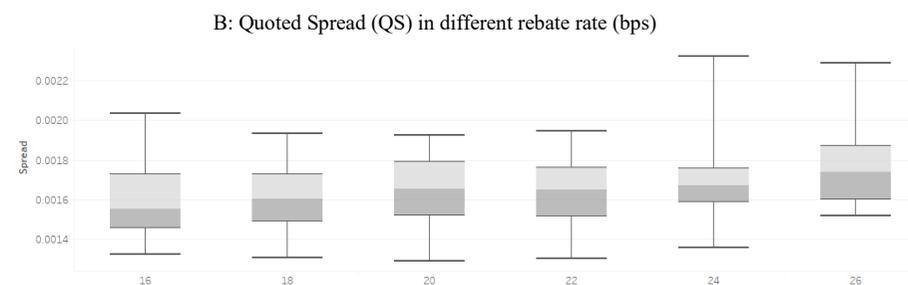

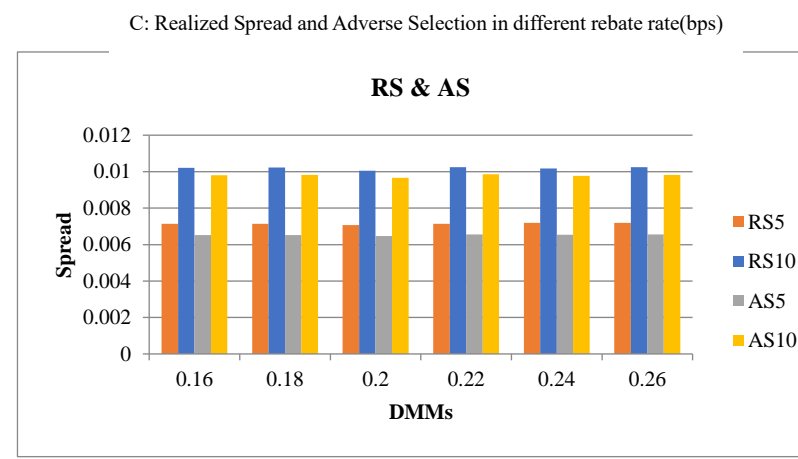

# Result Analysis

### Different Number of DMMs

The trials examining liquidity with incremental numbers of DMMs from 1 to 5 at a 20 bps rebate rate show a general improvement in market conditions. As the number of DMMs increases, the quote spread decreases, indicating better market liquidity due to heightened competition among DMMs. This competition compels DMMs to offer more favorable bid and ask prices to fulfill market orders. Statistical analysis using an ANOVA test (figure 5) confirms this observation, with an F-statistic of 2.68 and a P-value of 0.03388, which is significant at the alpha level of 0.05. Thus, we reject the null hypothesis that group means are equal, affirming that an increased number of DMMs negatively affects the quote spread. Additionally, market depth increases with more DMMs, enhancing the market's capacity to absorb larger orders without substantial price shifts. The significance of these differences is also supported by the ANOVA test (figure 6), indicating changes in market depth as the number of DMMs varies.

As the number of Designated Market Makers (DMMs) increases, the realized spread and adverse selection both show a downward trend. ANOVA results (figures 7-8) indicate a significant impact of the number of DMMs on the 5-minute and 10-minute Realized Spread (RS5 and RS10), with average values decreasing as more DMMs are added. This suggests improved liquidity and efficiency, as narrower spreads are observed within a 5 or 10-minute window after trade execution. The very low p-value enhances confidence in these results.

Similarly, the ANOVA test for 5-minute and 10-minute Adverse Selection (AS5 and AS10) in figures 9-10 shows a statistically significant difference across varying numbers of DMMs. A decrease in adverse selection risk as more DMMs join the market is indicated by the declining trend in average values, supported by an extremely low p-value.

Contrary to expectations, the Execution to Pending Ratio (EP_ratio) decreases with more DMMs, attributed to increased orders from additional DMMs. This suggests that an increase in DMMs does not necessarily correlate with enhanced order execution efficiency. To accurately assess this, filtering out orders from DMMs might be necessary to measure the true impact on the ratio.

### Different Rebate Rate

The experimental trials exploring different bps reveal complex effects on liquidity as the rebate rate for DMMs changes. Notably, as the rebate rate increases, the Execution to Pending Ratio (EP_ratio) also increases, suggesting that higher rebates may motivate DMMs to enhance their execution rates, potentially by tightening their quote spreads or being more aggressive in order fulfillment.

ANOVA results (figure 11) confirm a statistically significant difference in the EP_ratio across varying rebate rates in the three DMM scenarios. The trend indicates that higher rebate rates correlate with improved execution ratios, supported by an extraordinarily low p-value, underscoring strong confidence in these findings.

Conversely, the quoted spread data, as shown in both the graph and the ANOVA test (figure 12), present an unexpected trend: the average quoted spread (QS) decreases as the rebate rate decreases. This result suggests that the incentive structure may be influencing DMMs' behaviors in ways not initially anticipated. Despite higher rebates intended to reduce quoted spreads, the significant results with a p-value below the 0.05 threshold indicate a contrary effect, challenging expectations about the influence of rebate incentives on DMM performance.

Analysis of the depth at the best quote (DB) across different rebate rate scenarios for three DMMs shows no statistically significant differences (figure 13). This indicates that, within the tested range, changes in rebate rates do not significantly impact the market depth provided by the DMMs.

Furthermore, there is no consistent trend in the 5-minute and 10-minute Realized Spreads (RS5 and RS10); they fluctuate with changes in the rebate rate. The 5-minute and 10-minute Adverse Selection (AS5 and AS10) also show slight fluctuations but generally increase as the rebate rate decreases.

Overall, while increasing the rebate rate has a generally positive effect on liquidity as reflected in the EP_ratio, its impact on the quoted spread and adverse selection presents more complexity. Contrary to expectations, higher rebate rates do not lead to lower quoted spreads or resolve information asymmetry issues.

**Figure 3: Sensitivity Matrix of Quoted Spread of 1-5 DMMs With Rebate Rate 0.16 – 0.26**

The quoted spreads in sensitivity matrix are the calculated by adjusting the number of DMMs and rebate rate and take the mean value of each scenario. The numbers are then converted to z-scores. Overall, optimal liquidity conditions—indicated by lower z-scores or tighter spreads—do not consistently align with higher rebate rates. Instead, a complex interaction suggests that mid-range rebate rates might offer better liquidity, particularly in scenarios involving more DMMs.

|      | 1MM      | 2MM      | 3MM      | 4MM      | 5MM      |
|------|----------|----------|----------|----------|----------|
| 0.16 | -0.22909 | -0.82016 | -0.55775 | -1.65052 | -1.08781 |
| 0.18 | 1.088755 | -0.78255 | -0.27386 | -1.65052 | -2.01994 |
| 0.2  | 0.948233 | -0.35327 | 0.145351 | -1.07563 | -0.92809 |
| 0.22 | 1.375381 | 0.819914 | 0.139185 | -0.20262 | -0.27478 |
| 0.24 | 1.713711 | 0.928039 | 0.736798 | 0.29836  | -0.79744 |
| 0.26 | 1.549176 | 0.33795  | 1.51958  | 0.794343 | 0.309275 |

**Figure 4: Price Efficiency across various number of DMMs**

The price efficiency (PE) is measured by

$$PE = \frac{abs(market\_midprice - fundamental\_value)}{fundamental\_value}$$

Whereas market_midprice is measured as

$$market\_midprice = \frac{(best\_ask + best\_bid)}{2}$$

The introduction of a few DMMs initially brings market prices closer to fundamental values, as shown by lower PE scores, suggesting improved efficiency. However, as the number of DMMs increases, PE scores do not continue to decline but plateau, indicating that there may be an optimal number of DMMs—possibly around 1 or 7—that maximizes price efficiency. Beyond this point, additional DMMs do not appear to offer further benefits, and high PE scores with a large number of DMMs suggest that excessive DMM participation could disrupt pricing decisions.

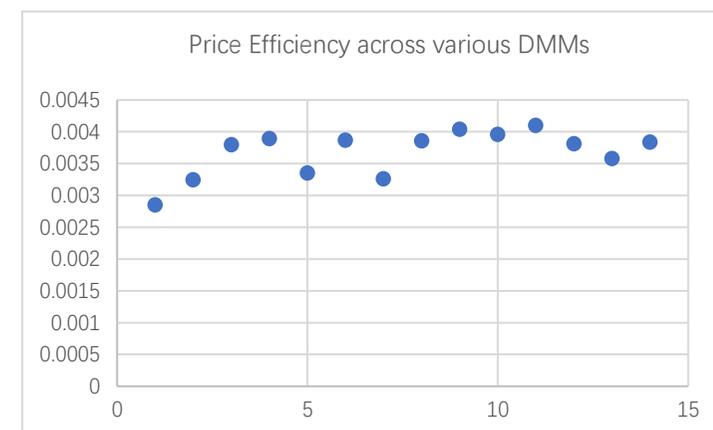

**Figure 17: DMM Performance in Different Number of DMMs Setting**

We leveraged the existing framework to analyze the performance of Designated Market Makers (DMMs) under different scenarios involving varying numbers of DMMs. In scenarios with multiple DMMs, we randomly selected one DMM for evaluation. Consistent with other simulations, the results presented are based on the average of 30 simulations of each scenario.

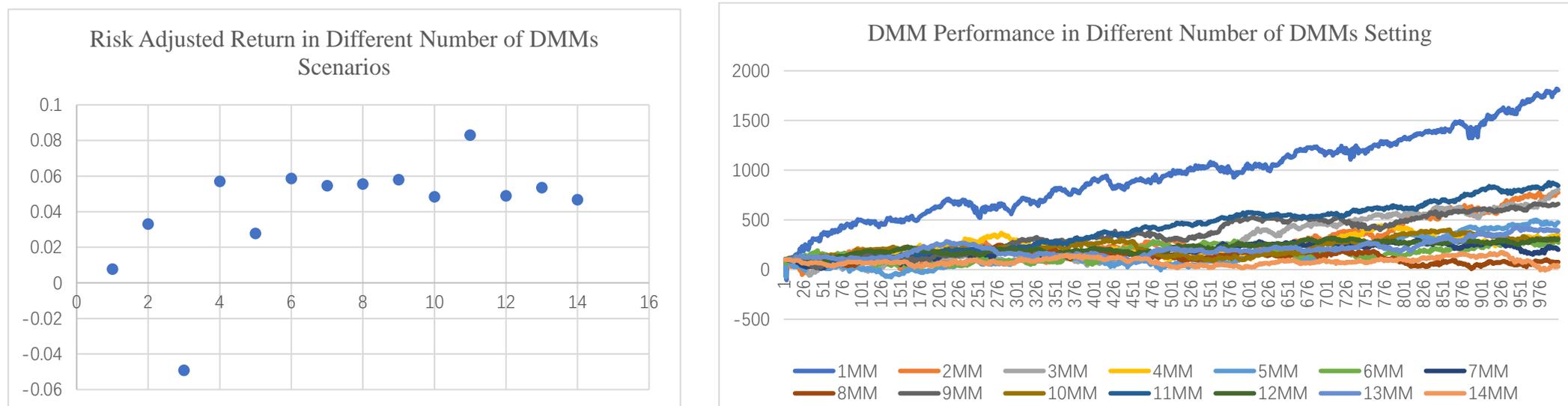

While we anticipated that increased competition in scenarios with more Designated Market Makers (DMMs) would result in lower profits for each individual DMM, the results did not support this assumption. Scenarios with a single DMM showed the highest absolute returns, but also the greatest volatility. Consequently, when returns are adjusted for risk, they are slightly lower than those in scenarios with multiple DMMs. This may be because fewer DMMs also bear greater risks from market shocks and informed traders. In scenarios with more participants, some DMMs may exploit opportunities to profit from their counterparts, thereby enhancing their profitability in a competitive environment.

**Sensitivity Matrix**

We can visualize the impact of varying the number of DMMs and rebate rates on market liquidity through a sensitivity heat map, as shown in Figure 3. This sensitivity matrix, displaying z-scores of quoted spreads, reveals that higher rebate rates generally correlate with wider spreads across all DMM configurations, suggesting poorer liquidity. This might indicate that higher rebates are not effectively motivating DMMs to narrow their spreads or that they allow DMMs to maintain wider spreads while still attracting trades.

Despite this, the EP_ratio increases with the number of DMMs, implying that DMMs are fulfilling more orders near the midprice, leading to larger spreads in the remaining limited orders at each timestamp. This could indicate a flaw in the system design as we divide the market into discrete timestamps. If all participants make decisions at the same time, the quoted spread after the exchange matches orders may not be very meaningful, as the best bid and ask may have been consumed.

The response to increasing rebate rates varies with the number of DMMs. With one market maker (1MM), z-scores begin low at the smallest rebate rate, increase, then decrease, suggesting a non-linear relationship between rebate rates and liquidity. Scenarios with four and five market makers (4MM and 5MM) show a general trend of increasing z-scores with rising rebate rates, yet liquidity seems less affected compared to scenarios with two and three market makers (2MM and 3MM).

Overall, optimal liquidity conditions—indicated by lower z-scores or tighter spreads—do not consistently align with higher rebate rates. Instead, a complex interaction suggests that mid-range rebate rates might offer better liquidity, particularly in scenarios involving more DMMs.

**Price Efficiency**

The price efficiency chart assesses the impact of different numbers of DMMs on market pricing. Higher Price Efficiency (PE) scores, indicating a greater deviation from the fundamental value, suggest less efficient pricing.

Contrary to expectations that more DMMs would enhance price efficiency, the data reveals a nuanced story. The introduction of a few DMMs initially brings market prices closer to fundamental values, as shown by lower PE scores, suggesting improved efficiency. However, as the number of DMMs increases, PE scores do not continue to decline but plateau, indicating that there may be an optimal number of DMMs—possibly around 1 or 7—that maximizes price efficiency. Beyond this point, additional DMMs do not appear to offer further benefits, and high PE scores with a large number of DMMs suggest that excessive DMM participation could disrupt pricing decisions.

While a single DMM may offer better price efficiency, its impact on market liquidity is less favorable. Thus, finding an optimal range of DMMs is crucial, one that balances enhanced market liquidity with maintained price efficiency.

**Figure 15: Market Reaction to Shocks in Different Number of DMMs**

The study examined the effect of different numbers of DMMs on the market's ability to correct prices following a 30% increase in stock fundamentals. We measure based on the timestamps required for market midprice to converge within 1% deviation of the fundamental price.

SUMMARY

| Groups | Count | Sum | Average | Variance |
|--------|-------|-----|---------|----------|
| MM1 | 31 | 392 | 12.64516 | 105.0366 |
| MM2 | 31 | 439 | 14.16129 | 111.0731 |
| MM3 | 31 | 296 | 9.548387 | 58.85591 |
| MM4 | 31 | 348 | 11.22581 | 106.0473 |
| MM5 | 31 | 447 | 14.41935 | 146.4516 |
| MM6 | 31 | 387 | 12.48387 | 148.1247 |
| MM7 | 31 | 500 | 16.12903 | 280.5161 |
| MM8 | 31 | 364 | 11.74194 | 143.9978 |
| MM9 | 31 | 476 | 15.35484 | 265.9032 |
| MM10 | 31 | 611 | 19.70968 | 662.8129 |

This unexpected outcome may be attributed to increased uncertainty in the market due to the presence of more participants. Scenarios with more DMMs displayed significantly larger variances, suggesting that higher activity levels can lead to increased order spreads and added market noise, making trends harder to discern. The scenario of 3 DMMs having the lowest average and variance may imply the existence of an optimal setting that maximizes the efficiency of price correction without adding too much market noise. Consequently, an excessive number of DMMs does not necessarily enhance market correction efficiency and can instead obscure market trends.

**Figure 16: Simulation Result of Liquidity in 3 DMMs, 20 bps Rebate Rate and Different Traded Stocks**

This series evaluates market liquidity with three DMM, 20 bps rebate rate and three different trading assets. We chose stocks from three publicly traded companies: Coca-Cola (NYSE: KO), Starbucks (Nasdaq: SBUX), and Nvidia (Nasdaq: NVDA). These companies exhibited significant differences in returns and volatility over the past year. We simulate the market with different traded stocks by varying the return and volatility of the asset.

| Stock | Annual Return | Annual Volatility |
|-------|---------------|-------------------|
| KO    | 0.0236        | 0.13              |
| SBUX  | -0.27         | 0.254             |
| NVDA  | 2.07          | 0.495             |

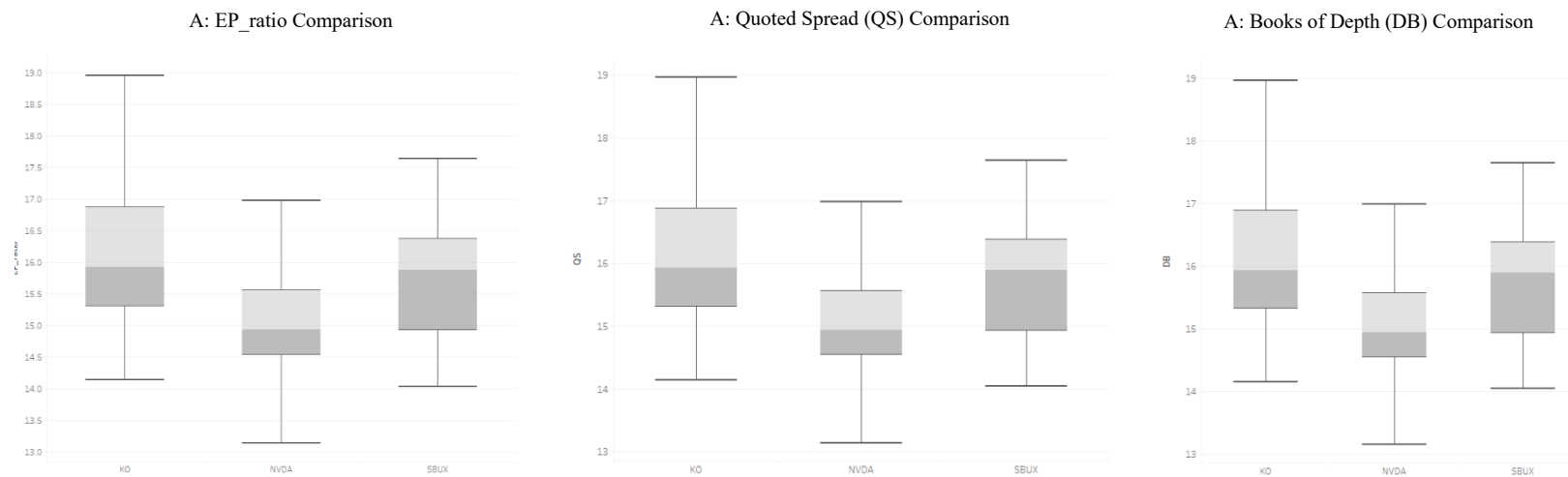

The results demonstrate a highly significant difference in market liquidity. As indicated by the EP_ratio, Books of Depth, Realized Spread and Adverse Selection among Coca-Cola, NVIDIA, and Starbucks. Coca-Cola shows the highest liquidity, followed by Starbucks and then NVIDIA. The quoted spread shows the exact opposite result of our expectation possibly due to the same system flaw discussed in "Different Rebate Rate" section.

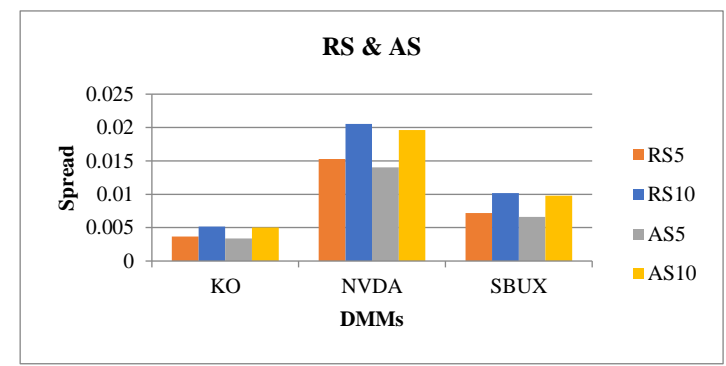

**Reaction to Shocks**

The study examined the effect of different numbers of DMMs on the market's ability to correct prices following a 30% increase in stock fundamentals (Figure 14). The results show a wide distribution in the number of days required for price correction, with no clear difference across scenarios involving varying numbers of DMMs. Furthermore, the ANOVA test did not reveal any significant differences, contradicting expectations that more DMMs, by providing greater liquidity and market depth, would reduce correction times.

This unexpected outcome may be attributed to increased uncertainty in the market due to the presence of more participants. Scenarios with more DMMs displayed significantly larger variances, suggesting that higher activity levels can lead to increased order spreads and added market noise, making trends harder to discern. The scenario of 3 DMMs having the lowest average and variance may imply the existence of an optimal setting that maximizes the efficiency of price correction without adding too much market noise. Consequently, an excessive number of DMMs does not necessarily enhance market correction efficiency and can instead obscure market trends.

**Liquidity in different assets**

Given varying returns and volatilities, market liquidity demonstrates significant differences across assets. Coca-Cola (KO) showcases the highest average EP_ratio, the deepest book of depth, and the lowest realized spread and adverse selection, positioning it as the most liquid market among the three assessed assets. In contrast, Nvidia features the least liquid market. Intriguingly, the quoted spread contradicts our expectations, possibly due to a systemic flaw in measuring unsettled orders, as discussed in the "Different Rebate Rate" section.

This simulation solely manipulates the stocks' returns and volatilities. In reality, trading volume, which varies substantially across different stocks, is a primary driver of market liquidity. Consequently, while the simulation results, supported by the ANOVA test, indicate that different assets can significantly influence market liquidity through returns and volatilities, further research is necessary to understand the impacts of nuanced changes in these factors on market liquidity.

# Conclusion

Our study confirms that Designated Market Makers (DMMs) substantially enhance liquidity and contribute to narrower bid-ask spreads, better depth at the best quote, narrower realized spreads, and improved robustness against information asymmetry. However, our findings on the impact of DMMs on market response to shocks indicate no significant effect on the speed of price corrections after the market shocks, suggesting that the number of DMMs alone may not necessarily improve market correction efficiency.

Rebate rates for DMMs also exhibit a nuanced relationship with liquidity. While a better rebate rate helps fulfill more limited orders, surprisingly, higher rebate rates correspond with wider spreads, suggesting a counterintuitive effect where expected incentives do not align with liquidity enhancement. This conclusion might not be accurate due to an inherent flaw in the system's timestamp design, which could lead to incorrect measurements. Additional validation is necessary to confirm these findings. Despite speculation that orders with lower spreads might have been fulfilled first, the results indicate minimal effects of rebate rates on depth at the best quote, realized spreads, and adverse selection.

The variation in assets suggests that exchanges need to appropriately allocate, and likely vary, the number of Designated Market Makers (DMMs) to enhance market liquidity. Further analysis of DMMs' profit and loss performance indicates that increased market competition does not necessarily harm the DMM's performance as initially expected.

The sensitivity matrix further indicates that increasing rebate rates may not effectively enhance market liquidity and could lead to wider spreads. Moreover, the presence of more DMMs does not uniformly translate into better liquidity under higher rebate rates. This suggests an optimal point for rebate rates that balances the need to incentivize DMMs without adversely affecting the market liquidity they are supposed to foster.

While more DMMs can improve liquidity, too many DMMs can lead to price inefficiency, as their orders significantly affect market pricing. Conversely, too few DMMs may also result in inefficiency due to possibly insufficient liquidity. It would be advisable for exchanges or regulatory bodies to identify this optimal number of DMMs to ensure that market resources are used effectively without over-saturating the market with liquidity providers.

In light of these findings, we recommend that exchanges optimize the number of DMMs to harness the benefits of enhanced liquidity and revisit the incentive frameworks to better align with market efficiency goals. The insights gained herein pave the way for more targeted research into the calibration of DMM incentives and the broader implications for diverse financial instruments and market conditions.

# Future Direction

The findings from this study open several pathways for further investigation. It would be particularly beneficial for exchanges to explore the underlying reasons why higher rebate rates do not effectively enhance market liquidity. Gaining a clearer understanding could facilitate the development of more effective incentives for designated market makers to improve liquidity conditions. We will also further improve the simulation system to solve the possible quoted spread measurement problem. Additionally, the impact of different types of financial instruments on DMM performance requires further exploration. Our current research is limited to equities; therefore, examining how DMMs influence market liquidity in other asset classes such as options, futures, or cryptocurrencies could provide valuable insights into the broader applicability of our findings. Such studies could reveal significant differences in how liquidity is managed across various market structures and help tailor DMM strategies to specific financial environments.

# Appendix

Here's the statistical description of the liquidity metrics in 20 bps and different number (1-5) of DMM. Please refer to the methodology section for liquidity metrics.

Single DMM Scenario

|       | EP_ratio | qs       | DB        | RS5      | RS10     | AS5      | AS10     |
|-------|----------|----------|-----------|----------|----------|----------|----------|
| count | 30       | 30       | 30        | 28473    | 28473    | 28473    | 28473    |
| mean  | 0.505962 | 0.001712 | 14.6135   | 0.00749  | 0.010489 | 0.006728 | 0.009965 |
| std   | 0.025675 | 0.000149 | 0.608753  | 0.006221 | 0.008419 | 0.005669 | 0.008008 |
| min   | 0.472967 | 0.001426 | 13.636    | 0        | 0        | 0        | 0        |
| 25%   | 0.487294 | 0.001611 | 14.134    | 0.002743 | 0.004027 | 0.002428 | 0.003798 |
| 50%   | 0.5042   | 0.001707 | 14.576    | 0.005963 | 0.008484 | 0.005328 | 0.008064 |
| 75%   | 0.516521 | 0.001835 | 14.979    | 0.01066  | 0.01493  | 0.009485 | 0.014193 |

Two DMMs Scenario

|       | EP_ratio | qs       | DB        | RS5      | RS10     | AS5      | AS10     |
|-------|----------|----------|-----------|----------|----------|----------|----------|
| count | 30       | 30       | 30        | 28240    | 28240    | 28240    | 28240    |
| mean  | 0.417412 | 0.001615 | 15.3337   | 0.007264 | 0.01035  | 0.006587 | 0.009861 |
| std   | 0.022992 | 0.000126 | 1.072425  | 0.00604  | 0.008332 | 0.005555 | 0.007982 |
| min   | 0.36095  | 0.001401 | 13.433    | 0        | 0        | 0        | 0        |
| 25%   | 0.403373 | 0.0015   | 14.609    | 0.002685 | 0.003973 | 0.002399 | 0.003744 |
| 50%   | 0.420185 | 0.001623 | 15.201    | 0.005793 | 0.008405 | 0.005227 | 0.008024 |
| 75%   | 0.430847 | 0.001718 | 16.195    | 0.010242 | 0.014617 | 0.009256 | 0.013951 |

Three DMMs Scenario

|       | EP_ratio | qs       | DB        | RS5      | RS10     | AS5      | AS10     |
|-------|----------|----------|-----------|----------|----------|----------|----------|
| count | 30       | 30       | 30        | 28087    | 28087    | 28087    | 28087    |
| mean  | 0.370293 | 0.001652 | 15.92717  | 0.007071 | 0.010044 | 0.006473 | 0.00966  |
| std   | 0.024789 | 0.000156 | 0.867236  | 0.006009 | 0.008377 | 0.005594 | 0.008044 |
| min   | 0.322326 | 0.001293 | 14.209    | 0        | 0        | 0        | 0        |
| 25%   | 0.351399 | 0.001522 | 15.34225  | 0.00262  | 0.003751 | 0.002346 | 0.003665 |
| 50%   | 0.371786 | 0.001655 | 15.783    | 0.005566 | 0.007985 | 0.005069 | 0.007685 |
| 75%   | 0.388792 | 0.001777 | 16.56025  | 0.009899 | 0.014178 | 0.009025 | 0.013526 |

Four DMMs Scenario

|       | EP_ratio | qs       | DB        | RS5      | RS10     | AS5      | AS10     |
|-------|----------|----------|-----------|----------|----------|----------|----------|
| count | 30       | 30       | 30        | 28053    | 28053    | 28053    | 28053    |
| mean  | 0.344156 | 0.001626 | 16.45207  | 0.007073 | 0.010058 | 0.006487 | 0.009662 |
| std   | 0.032619 | 0.000203 | 1.250949  | 0.006049 | 0.008421 | 0.005652 | 0.008103 |
| min   | 0.272332 | 0.001325 | 13.974    | 0        | 0        | 0        | 0        |
| 25%   | 0.327903 | 0.001467 | 15.69075  | 0.002622 | 0.003731 | 0.00233  | 0.003589 |
| 50%   | 0.342824 | 0.001611 | 16.3655   | 0.005563 | 0.008059 | 0.005085 | 0.007715 |
| 75%   | 0.359299 | 0.00174  | 17.23575  | 0.009877 | 0.014111 | 0.009036 | 0.013532 |

Five Market Makers Scenario

|  | EP_ratio | qs | DB | RS5 | RS10 | AS5 | AS10 |
|---|---|---|---|---|---|---|---|
| count | 30 | 30 | 30 | 28022 | 28022 | 28022 | 28022 |
| mean | 0.325937 | 0.001572 | 16.57463 | 0.006858 | 0.009762 | 0.00637 | 0.00946 |
| std | 0.027388 | 0.000212 | 1.257239 | 0.005866 | 0.008284 | 0.005456 | 0.007962 |
| min | 0.267979 | 0.001257 | 14.089 | 0 | 0 | 0 | 0 |
| 25% | 0.301364 | 0.001432 | 15.88775 | 0.002495 | 0.00356 | 0.002321 | 0.003479 |
| 50% | 0.328774 | 0.001541 | 16.577 | 0.005412 | 0.007705 | 0.00503 | 0.007536 |
| 75% | 0.347216 | 0.001638 | 17.5225 | 0.009629 | 0.013766 | 0.008882 | 0.013285 |

On the other aspect, we maintain 3 designated market makers and iterate the rebate rate from 16 bps to 26 bps.

|  | EP_ratio | qs | DB | RS5 | RS10 | AS5 | AS10 |
|---|---|---|---|---|---|---|---|
| -0.16 | 0.316113 | 0.001568 | 15.627 | 0.00705 | 0.010023 | 0.006474 | 0.009618 |
| -0.18 | 0.346223 | 0.00155 | 16.1248 | 0.006927 | 0.009825 | 0.006371 | 0.00948 |
| -0.2 | 0.368147 | 0.001633 | 16.0081 | 0.007175 | 0.010184 | 0.006602 | 0.009845 |
| -0.22 | 0.402989 | 0.001585 | 16.18807 | 0.006988 | 0.009895 | 0.00637 | 0.009494 |
| -0.24 | 0.436579 | 0.001715 | 16.30627 | 0.007097 | 0.010015 | 0.006472 | 0.00962 |

**Figure 5: Quoted Spread (QS) in 20 bps rebate rate**

SUMMARY

| Groups | Count | Sum | Average | Variance |
|---|---|---|---|---|
| 1MM | 30 | 0.051345 | 0.001712 | 2.23E-08 |
| 2MM | 30 | 0.048437 | 0.001615 | 1.59E-08 |
| 3MM | 30 | 0.049551 | 0.001652 | 2.45E-08 |
| 4MM | 30 | 0.048773 | 0.001626 | 4.11E-08 |
| 5MM | 30 | 0.047152 | 0.001572 | 4.5E-08 |

ANOVA

| Source of Variation | SS | df | MS | F | P-value | F crit |
|---|---|---|---|---|---|---|
| Between Groups | 3.19E-07 | 4 | 7.98E-08 | 2.68243 | 0.033888 | 2.434065 |
| Within Groups | 4.31E-06 | 145 | 2.97E-08 | | | |
| Total | 4.63E-06 | 149 | | | | |

A p-value of 0.033, which is less than the significance threshold of 0.05, indicates that variations in the number of Designated Market Makers (DMMs) significantly affect the quoted spread.

**Figure 6: DB (Depth of the Orderbook) in 20 bps rebate rate**

SUMMARY

| Groups | Count | Sum | Average | Variance |
|--------|-------|-----|---------|----------|
| 1MM | 30 | 438.405 | 14.6135 | 0.37058 |
| 2MM | 30 | 460.011 | 15.3337 | 1.150096 |
| 3MM | 30 | 477.815 | 15.92717 | 0.752098 |
| 4MM | 30 | 493.562 | 16.45207 | 1.564872 |
| 5MM | 30 | 497.239 | 16.57463 | 1.58065 |

ANOVA

| Source of Variation | SS | df | MS | F | P-value | F crit |
|---------------------|-----|-----|-----|-----|---------|--------|
| Between Groups | 79.94038 | 4 | 19.9851 | 18.44223 | 2.86E-12 | 2.434065 |
| Within Groups | 157.1306 | 145 | 1.083659 | | | |
| | | | | | | |
| Total | 237.071 | 149 | | | | |

A p-value of approximate 0, which is less than the significance threshold of 0.05, indicates that variations in the number of Designated Market Makers (DMMs) significantly affect the depth of the orderbook.

**Figure 7: RS5(5-min Realized Spread) in 20 bps rebate rate**

SUMMARY

| Groups | Count | Sum | Average | Variance |
|---|---|---|---|---|
| 1MM | 28473 | 213.2735 | 0.00749 | 3.87E-05 |
| 2MM | 28240 | 205.132 | 0.007264 | 3.65E-05 |
| 3MM | 28087 | 198.6067 | 0.007071 | 3.61E-05 |
| 4MM | 28053 | 198.4189 | 0.007073 | 3.66E-05 |
| 5MM | 28022 | 192.1794 | 0.006858 | 3.44E-05 |

ANOVA

| Source of Variation | SS | df | MS | F | P-value | F crit |
|---|---|---|---|---|---|---|
| Between Groups | 0.006392 | 4 | 0.001598 | 43.82118 | 8.09E-37 | 2.371995 |
| Within Groups | 5.136886 | 140870 | 3.65E-05 | | | |
| Total | 5.143278 | 140874 | | | | |

A p-value of approximate 0, which is less than the significance threshold of 0.05, indicates that variations in the number of Designated Market Makers (DMMs) significantly affect the 5-min Realized Spread.

**Figure 8: RS10(10-min Realized Spread) in 20 bps rebate rate**

SUMMARY

| Groups | Count | Sum | Average | Variance |
|--------|-------|-----|---------|----------|
| 1MM | 28473 | 298.667 | 0.010489 | 7.09E-05 |
| 2MM | 28240 | 292.2714 | 0.01035 | 6.94E-05 |
| 3MM | 28087 | 282.1084 | 0.010044 | 7.02E-05 |
| 4MM | 28053 | 282.1498 | 0.010058 | 7.09E-05 |
| 5MM | 28022 | 273.5599 | 0.009762 | 6.86E-05 |

ANOVA

| Source of Variation | SS | df | MS | F | P-value | F crit |
|---------------------|-----|-----|-----|-----|---------|--------|
| Between Groups | 0.009162 | 4 | 0.002291 | 32.71976 | 2.6E-27 | 2.371995 |
| Within Groups | 9.861708 | 140870 | 7E-05 | | | |
| | | | | | | |
| Total | 9.87087 | 140874 | | | | |

A p-value of approximate 0, which is less than the significance threshold of 0.05, indicates that variations in the number of Designated Market Makers (DMMs) significantly affect the 10-min Realized Spread.

**Figure 9: AS5(5-min adverse selection) in 20 bps rebate rate**

SUMMARY

| Groups | Count | Sum | Average | Variance |
|--------|-------|-----|---------|----------|
| 1MM | 28473 | 191.5664 | 0.006728 | 3.21E-05 |
| 2MM | 28240 | 186.0093 | 0.006587 | 3.09E-05 |
| 3MM | 28087 | 181.8009 | 0.006473 | 3.13E-05 |
| 4MM | 28053 | 181.9701 | 0.006487 | 3.2E-05 |
| 5MM | 28022 | 178.5124 | 0.00637 | 2.98E-05 |

ANOVA

| Source of Variation | SS | df | MS | F | P-value | F crit |
|---------------------|-----|-----|-----|-----|---------|--------|
| Between Groups | 0.002065 | 4 | 0.000516 | 16.54591 | 1.46E-13 | 2.371995 |
| Within Groups | 4.395832 | 140870 | 3.12E-05 | | | |
| | | | | | | |
| Total | 4.397898 | 140874 | | | | |

A p-value of approximate 0, which is less than the significance threshold of 0.05, indicates that variations in the number of Designated Market Makers (DMMs) significantly affect the 5-min adverse selection.

**Figure 10: AS10(10min adverse selection) in 20 bps rebate rate**

SUMMARY

| Groups | Count | Sum | Average | Variance |
|--------|-------|-----|---------|----------|
| 1MM | 28473 | 283.7272 | 0.009965 | 6.41E-05 |
| 2MM | 28240 | 278.4855 | 0.009861 | 6.37E-05 |
| 3MM | 28087 | 271.3074 | 0.00966 | 6.47E-05 |
| 4MM | 28053 | 271.062 | 0.009662 | 6.57E-05 |
| 5MM | 28022 | 265.0976 | 0.00946 | 6.34E-05 |

ANOVA

| Source of Variation | SS | df | MS | F | P-value | F crit |
|---------------------|-----|-----|-----|-----|---------|--------|
| Between Groups | 0.004354 | 4 | 0.001089 | 16.92434 | 7E-14 | 2.371995 |
| Within Groups | 9.060966 | 140870 | 6.43E-05 | | | |
| Total | 9.06532 | 140874 | | | | |

A p-value of approximate 0, which is less than the significance threshold of 0.05, indicates that variations in the number of Designated Market Makers (DMMs) significantly affect the 10-min adverse selection.

**Figure 11: EP_ratio (Executed/Pending orders) in 3DMMs**

SUMMARY

| Groups | Count | Sum | Average | Variance |
|---|---|---|---|---|
| 26 | 30 | 13.79699 | 0.4599 | 0.001043 |
| 24 | 30 | 12.98134 | 0.432711 | 0.000714 |
| 22 | 30 | 11.94835 | 0.398278 | 0.001029 |
| 20 | 30 | 11.10879 | 0.370293 | 0.000615 |
| 18 | 30 | 10.14475 | 0.338158 | 0.000577 |
| 16 | 30 | 9.482485 | 0.316083 | 0.000393 |

ANOVA

| Source of Variation | SS | df | MS | F | P-value | F crit |
|---|---|---|---|---|---|---|
| Between Groups | 0.456533 | 5 | 0.091307 | 125.3621 | 9.08E-56 | 2.266062 |
| Within Groups | 0.126732 | 174 | 0.000728 | | | |
| | | | | | | |
| Total | 0.583265 | 179 | | | | |

A p-value of approximate 0, which is less than the significance threshold of 0.05, indicates that variations in rebate rate significantly affect the EP_ratio (Executed/Pending orders).

**Figure 12: Quoted Spread (QS) in 3DMMs**

SUMMARY

| Groups | Count | Sum | Average | Variance |
|---|---|---|---|---|
| 26 | 30 | 0.052622 | 0.001754 | 3.47E-08 |
| 24 | 30 | 0.050873 | 0.001696 | 4.3E-08 |
| 22 | 30 | 0.049537 | 0.001651 | 2.96E-08 |
| 20 | 30 | 0.049551 | 0.001652 | 2.45E-08 |
| 18 | 30 | 0.048614 | 0.00162 | 3.09E-08 |
| 16 | 30 | 0.04798 | 0.001599 | 3.09E-08 |

ANOVA

| Source of Variation | SS | df | MS | F | P-value | F crit |
|---|---|---|---|---|---|---|
| Between Groups | 4.65E-07 | 5 | 9.3E-08 | 2.879925 | 0.015912 | 2.266062 |
| Within Groups | 5.62E-06 | 174 | 3.23E-08 | | | |
| Total | 6.08E-06 | 179 | | | | |

A p-value of 0.016, which is less than the significance threshold of 0.05, indicates that variations in rate significantly affect the Quoted Spread.

**Figure 13: DB (Depth of the Orderbook) in 3DMMs**

SUMMARY

| Groups | Count | Sum | Average | Variance |
|---|---|---|---|---|
| 26 | 30 | 483.457 | 16.11523 | 0.731405 |
| 24 | 30 | 477.6 | 15.92 | 0.832124 |
| 22 | 30 | 489.319 | 16.31063 | 0.898681 |
| 20 | 30 | 477.815 | 15.92717 | 0.752098 |
| 18 | 30 | 479.134 | 15.97113 | 1.16491 |
| 16 | 30 | 470.136 | 15.6712 | 1.148509 |

ANOVA

| Source of Variation | SS | df | MS | F | P-value | F crit |
|---|---|---|---|---|---|---|
| Between Groups | 6.876762 | 5 | 1.375352 | 1.492858 | 0.194465 | 2.266062 |
| Within Groups | 160.3041 | 174 | 0.921288 | | | |
| | | | | | | |
| Total | 167.1809 | 179 | | | | |

A p-value of 0.194465, which is larger than the significance threshold of 0.05, does not indicate that variations in rebate rate significantly affect the depth of the orderbook.

**Figure 14: Market Reaction to Shocks**

SUMMARY

| Groups | Count | Sum | Average | Variance |
|---|---|---|---|---|
| MM1 | 31 | 392 | 12.64516 | 105.0366 |
| MM2 | 31 | 439 | 14.16129 | 111.0731 |
| MM3 | 31 | 296 | 9.548387 | 58.85591 |
| MM4 | 31 | 348 | 11.22581 | 106.0473 |
| MM5 | 31 | 447 | 14.41935 | 146.4516 |
| MM6 | 31 | 387 | 12.48387 | 148.1247 |
| MM7 | 31 | 500 | 16.12903 | 280.5161 |
| MM8 | 31 | 364 | 11.74194 | 143.9978 |
| MM9 | 31 | 476 | 15.35484 | 265.9032 |
| MM10 | 31 | 611 | 19.70968 | 662.8129 |

ANOVA

| Source of Variation | SS | df | MS | F | P-value | F crit |
|---|---|---|---|---|---|---|
| Between Groups | 2332.774 | 9 | 259.1971 | 1.277576 | 0.248386 | 1.911151 |
| Within Groups | 60864.58 | 300 | 202.8819 | | | |
| | | | | | | |
| Total | 63197.35 | 309 | | | | |

A p-value of 0.248, which is larger than the significance threshold of 0.05, does not indicate that variations in the number of DDMs significantly affect the timestamps required for market correction after shocks.

Program output examples

Example of the cash and inventory report of all participants

|   | mm1_inventory | mm1_cash | p1_inventory | p1_cash | p3_inventory | p3_cash | price_history | fundamental | difference |
|---|---|---|---|---|---|---|---|---|---|
| 0 | 0 | 100 | 0 | 100 | 0 | 100 | 133.9 | 134 | 0.000746 |
| 1 | 0 | 100 | 5 | -570 | 0 | 100 | 133.9 | 133.90038 | 2.84E-06 |
| 2 | 50 | -6580.11 | 10 | -1237.5 | -1 | 233.9 | 133.92 | 133.315612 | 0.004534 |
| 3 | 122 | -16193.6 | 10 | -1237.5 | -11 | 1566.8 | 133.92 | 133.952649 | 0.000244 |

Example of the realized trades across all periods

|   | Time | Matched P | Matched Q | Buyer | Seller |
|---|---|---|---|---|---|
| 0 | 1 | 134 | 4 | -1 | 0 |
| 1 | 1 | 134 | 1 | -1 | 0 |
| 2 | 2 | 133.77 | 10 | 1 | 1 |

Example of all trades status at the end of each timestamp

|   | order_id | customer_i | order_time | asset | order_type | order_price | order_qua | total_amo | status |
|---|---|---|---|---|---|---|---|---|---|
| 0 | 1IBM35 | 0 | 1 | IBM | SELL | 134.1 | 8 | 1072.8 | PENDING |
| 1 | 1IBM99 | 0 | 1 | IBM | SELL | 134 | 0 | 536 | EXECUTED |
| 2 | 1IBM23 | 0 | 1 | IBM | SELL | 134 | 4 | 670 | PENDING |
| 3 | 1IBM2 | 0 | 1 | IBM | SELL | 134 | 6 | 804 | PENDING |